# Bipartisan politics and poverty as a risk factor for contagion and mortality from SARS-CoV-2 virus in the United States of America

Dr. Cesar R. Salas-Guerra

*EDUCTUM Business School, Miami, United States of America*

*UAB, Universitat Autónoma de Barcelona, Spain*

*cesar.salas@eductum.us*

*cesar.salasg@e-campus-uab.cat*

*ORCID ID: 0000-0001-7182-3002*

*https://doi.org/10.13140/RG.2.2.36653.61925*

# Abstract

In the United States, from the start of the COVID-19 pandemic to December 31, 2020, 341,199 deaths and more than 19,663,976 infections were recorded. Recent literature establishes that communities with poverty-related health problems, such as obesity, cardiovascular disease, diabetes, and hypertension, are more susceptible to mortality from SARS-CoV-2 infection. Additionally, controversial public health policies implemented by the nation's political leaders have highlighted the socioeconomic inequalities of minorities. Therefore, through multivariate correlational analysis using machine learning techniques and structural equations, we measure whether social determinants are associated with increased infection and death from COVID-19 disease. The PLS least squares regression analysis allowed identifying a significant impact between social determinants and COVID-19 disease through a predictive value of $R2 = .916$, $\beta = .836$, $p =. 000$ (t-value = 66,137) shows that for each unit of increase in social determinants, COVID-19 disease increases by 91.6%. The clustering index used for correlational analysis generated a new data set comprising three groups: C1 Republicans, C2 and C3 Democrats from California, New York, Texas, and Florida. This analysis made it possible to identify the poverty variable as the main risk factor related to the high rates of infection in Republican states and a high positive correlation between the population not insured with a medical plan and high levels of virus contagion in the states of group C3. These findings explain the argument that poverty and lack of economic security put the public or private health system at risk and calamity.

**Introduction**

Moral economics criticizes the contemporary economic model's current position, which establishes the predominance of capital over human well-being and the criteria for classifying ideological institutions considering some inferior human beings due to their race, skin color, or religion (Bolton & Laaser, 2013).

These inequalities reflect vulnerable communities characterized by their housing conditions, transportation, language barriers, population density, health conditions, and medical care access (Smittenaar, 2020). Unfortunately, this population is more vulnerable to increased exposure to SARS-CoV-2 contagion and specific health considerations such as obesity, diabetes, and hypertension (Patel et al., 2020).

From the beginning of the first infections until December 31, 2020, 341,199 people died in the United States, with more than 19,663,976 infected; this disease has a tremendous impact on medical care and the economy. The National Institute of Allergies and Infectious Diseases established guidelines that suggested the use of facial protection, social distancing, early diagnosis, and the respective follow-up of those infected as a strategy to contain transmission (Gremmels et al., 2020).

However, with great notoriety in the United States, the execution of public health policies to face the pandemic has been limited to ideological confrontations of the American bipartisanship, with deep inequalities that threaten the sense of justice, equity, and morality (Hadjisolomou & Simone, 2020).

Therefore, based on the problem raised above, we pose the following research question: Is poverty a risk factor attributed to the high infection rates and mortality from the SARS-CoV-2 virus? Next, we will describe the study variables.

**Classification and definition of the study variables**

This study's variables are made up of constructs and indicators, described below according to the order they were considered in the conceptual model.

Table 1. Variables definition matrix

| Code | Variable | Definition | References |
|---|---|---|---|
| **SDF** | **Social Determinants Factors** | They are socioeconomic variables that measure vulnerabilities related to people living below the poverty level, housing conditions, transportation, language barriers, population density, health conditions, and the population's access to medical care. | (Patel, 2020; Rolfe et al., 2020; Harrison et al., 2020; Smittenaar, 2020) |
| **C19-D** | **COVID-19 Disease** | SARS-CoV-2 is the severe acute respiratory syndrome caused by the (coronavirus 2) pathogen. | (Yao et al., 2020; Chen & Li, 2020) |

*Social Determinants Factors (SDF)*

- Covid-19 community vulnerability factors index (CCIV) measures how a community responds to the health, economic, and social consequences of virus. The indicators that make up this index are housing conditions, transportation, language barriers, population density, health conditions, and medical care access.

- Estimate number of poverty (POVER-20) poverty is defined as the lack of sufficient income to satisfy basic needs; This indicator measures the number of people living below the poverty level.

- Number of unemployed (UNEMP-20) measures the number of available people who stopped working and are looking for work. This indicator comprises unemployed persons who were available for work and made at least a specific and active effort to find a job during the survey's reference week; they were temporarily fired and expected to return to work. This indicator does not apply to sick people.
- Number of people without health insurance (UNINSU-20) measure the coverage rate for people without health insurance to help cover their medical expenses.
- Estimate number of population (POP-20) measure the population estimate for the year 2019.

*COVID-19 Disease (C19-D)*

- Contagions registered by Covid-19 (CS-COVID-19) measures the number of infections due to SARS-CoV-2 or severe acute respiratory syndrome.
- Death registered by Covid-19 (DE-COVID-19) measures the number of deaths from SARS-CoV-2 or severe acute respiratory syndrome.

**Theoretical framework**

*The economic nationalism and populist narrative*

Liberal economic theorists have faced the challenges of economic nationalists and trade protectionism that arise from the idea that economic activities should be subordinate to the nation's economic objectives (Helleiner, 2002). In recent years, with globalization, economic nationalism has changed so as not to disappear in the face of liberalism, reinforcing the motto of classic protectionism and liberal economic nationalism (Clift & Woll, 2012).

With the adoption of free trade, the development of new nationalist economic projects in America and Europe linked to political populism (Scheuerman, 2019) has led nations to move away from liberal economic policies through initiatives that promote different socioeconomic restrictions within the free trade, reinforcing the strategy of national autonomy through economic disintegration and deglobalization (Born et al., 2019).

This populist narrative characterized by a discursive framework based on different substantive ideological positions and an incoherent political agenda (Bonikowski, 2016) focuses on "economic nationalism" influenced by conservative proposals on issues of trade and international cooperation and immigration (Colantone & Stanig, 2019).

These conservative proposals and their populist narrative reinforced by the nationalists' militant individualism have led the United States of America to a profound deterioration of its political ideology under Donald Trump's presidency (Diaz & Mountz, 2020). The rapid spread of the COVID-19 pandemic and its effects on the exponential wave of infections and deaths has revealed the problems of inequality and poverty that affect thousands of American citizens.

### *The moral economy and social iniquity*

Poverty is defined as the lack of sufficient income to satisfy basic needs; in the United States, a person is poor when their income falls below a certain threshold of money, which is determined by the Census Bureau of the United States (Census, 2020). However, there is scientific evidence from studies that support the positive association between low income, low socioeconomic status, and low educational level with health conditions related to tobacco use, obesity, hypertension, cancer, and diabetes (Niessen et al., 2018).

The framework of the "moral economy" allows us to reflect on comprehensive organizational management policies and decisions where economic primacy is over human well-being, especially during economic crises such as COVID-19 produced by the severe acute respiratory syndrome or SARS-CoV-2 (Hadjisolomou & Simone, 2020). Current political agendas have generated insecurity for many marginalized minorities, which are part of a precarious labor system (Standing, 2016); the term "precarious" is known as a generalized state of insecurity that has tangible effects on the health of the individual (Harrison et al., 2020).

*The SARS-CoV-2 and social vulnerability*

SARS-CoV-2 is a severe acute respiratory syndrome caused by the pathogen (coronavirus 2) (Yao et al., 2020). This virus has a higher fatality rate among elderly patients and patients with comorbidities (Chen & Li, 2020). Within the symptoms, infected patients suffer from fever, dyspnoea, dry cough, pneumonia and fatigue accompanied by various non-respiratory clinical characteristics, such as gastrointestinal symptoms and eye inflammation (Hong et al., 2020).

According to studies carried out before the pandemic, socially and economically disadvantaged people are groups of greater vulnerability for developing health conditions (Cookson et al., 2016). Unfortunately, public policies have shown utter disregard for vulnerable groups, exposing thousands of human beings to mortality for decades (Marmot, 2005).

Unemployment is one of the biggest challenges in the COVID-19 pandemic, as the time of unemployment lengthens certain factors such as declining savings, and the limitations generated by unemployment insurance benefits wreak serious havoc in society (Chodorow-Reich, 2020).

Concerning COVID-19, among the risk factors identified in these disadvantaged groups is overcrowding in populated homes, which reduces compliance with social distancing standards, being employed in occupations that do not provide stable income or opportunities to work from home (Stewart, 2020).

*Social determinants factors*

Certain factors are critical in identifying the social determinants of health inequity, such as racism (Johnson, 2020), low-income households (Rolfe et al., 2020) and problems acquiring health plans (Weida et al., 2020). Therefore, the use of the COVID-19 Community Vulnerability Index (CCVI) will evaluate whether a community could respond to the health, economic and social consequences of COVID-19.

These socioeconomic indicators measure vulnerabilities related to housing conditions, transportation, language barriers, population density, health conditions, and the population's access to medical care (Stewart, 2020). The scientific evidence establishes that social disadvantage and vulnerability can influence the incidence of a health emergency similar to that of COVID-19 (Melvin et al., 2020). Consequently, poverty can not only increase exposure to the virus but also reduce the ability of the immune system to fight it, since people with low income are a negative determinant for access to medical care, this group being the highest risk mortality from COVID-19 (Patel et al., 2020). The discussion of the literature presented above allows us to propose the hypothesis of this research:

> H1: Poverty as a determining social factor drives infection and death from the SARS-CoV-2 virus disease.

*Methodological design of the research*

This research is quantitative - correlational, and it seeks to describe the relationship of the study variables at a given time (Sampieri & Lucio, 2013) using multivariate analysis statistics, clusters, and structural equations with partial least squares (Ajamieh, 2016) through the implementation of a state political control matrix to determine the impact relationships of social determinants in the COVID-19 disease.

The methodology was framed in the correlational-causal design because only the level of correlation between the variables was measured to identify possible causalities in the phenomenon that will later be studied (Orengo, 2008); The data used consisted of 408 observations structured in panel data obtained in the public repositories of the United States government that described below:

- The U.S. Census Bureau
- National Conference of State Legislatures
- Centers for Disease Control and Prevention, and
- Surgo Foundation Ventures

The data panel allowed to identify systematic and unobserved differences between the units correlated with factors whose effects should be measured (Wooldridge, 2009). Also, they allowed the results to be generalized since this study seeks to obtain from this population the data previously organized in tables methodologically designed for such purposes (Census, 2020).

**Political and geographical projection**

In the United States, there are two political parties, the Democrats and the Republicans (Rodden, 2010). However, as part of the study model, we identify and classify political parties that control state governments and decide public health (Ahler & Broockman, 2015).

This study covers 50 states and the District of Columbia, which is the main political and administrative unit (U.S. Department of Commerce, 2018). This distribution does not include American Samoa, Guam, the Northern Mariana Islands, Palau, Puerto Rico, and the United States Virgin Islands

Table 2. State party-control matrix

| Code | State | Party-Control | Code | State | Party-Control |
|---|---|---|---|---|---|
| 1 | Alabama | REPUBLICAN | 0 | Montana | DEMOCRATIC |
| 1 | Alaska | REPUBLICAN | 1 | Nebraska | REPUBLICAN |
| 1 | Arizona | REPUBLICAN | 0 | Nevada | DEMOCRATIC |
| 1 | Arkansas | REPUBLICAN | 1 | New Hampshire | REPUBLICAN |
| 0 | California | DEMOCRATIC | 0 | New Jersey | DEMOCRATIC |
| 0 | Colorado | DEMOCRATIC | 0 | New Mexico | DEMOCRATIC |
| 0 | Connecticut | DEMOCRATIC | 0 | New York | DEMOCRATIC |
| 0 | Delaware | DEMOCRATIC | 0 | North Caroline | DEMOCRATIC |
| 0 | District of Columbia | DEMOCRATIC | 1 | North Dakota | REPUBLICAN |
| 1 | Florida | REPUBLICAN | 1 | Ohio | REPUBLICAN |
| 1 | Georgia | REPUBLICAN | 1 | Oklahoma | REPUBLICAN |
| 0 | Hawaii | DEMOCRATIC | 1 | Oregon | REPUBLICAN |
| 1 | Idaho | REPUBLICAN | 0 | Pennsylvania | DEMOCRATIC |
| 0 | Illinois | DEMOCRATIC | 0 | Rhode Island | DEMOCRATIC |
| 1 | Indiana | REPUBLICAN | 1 | South Caroline | REPUBLICAN |
| 1 | Iowa | REPUBLICAN | 1 | South Dakota | REPUBLICAN |
| 0 | Kansas | DEMOCRATIC | 1 | Tennessee | REPUBLICAN |
| 0 | Kentucky | DEMOCRATIC | 1 | Texas | REPUBLICAN |
| 0 | Louisiana | DEMOCRATIC | 1 | Utah | REPUBLICAN |
| 0 | Maine | DEMOCRATIC | 1 | Vermont | REPUBLICAN |
| 1 | Maryland | REPUBLICAN | 0 | Virginia | DEMOCRATIC |
| 1 | Massachusetts | REPUBLICAN | 0 | Washington | DEMOCRATIC |
| 0 | Michigan | DEMOCRATIC | 1 | West Virginia | REPUBLICAN |
| 0 | Minnesota | DEMOCRATIC | 0 | Wisconsin | DEMOCRATIC |
| 1 | Mississippi | REPUBLICAN | 1 | Wyoming | REPUBLICAN |
| 1 | Missouri | REPUBLICAN | | | |

**Data Analysis**

In this study, 408 observations were analysed, organized into panel data; the process and tools are detailed below:

- The first analysis phase: reflective PLS model (Smart PLS 3.0)
- The second analysis phase: clustering and correlation analysis (Orange 3.0 learning machine platform)

*First analysis phase: reflective PLS model*

For this first phase, a non-parametric reflective model of partial least squares PLS and Bootstrapping is used since it is reliable and less sensitive to outliers. The model consists of two constructs, and fifty indicators previously explained.

Figure 1. Research empirical model

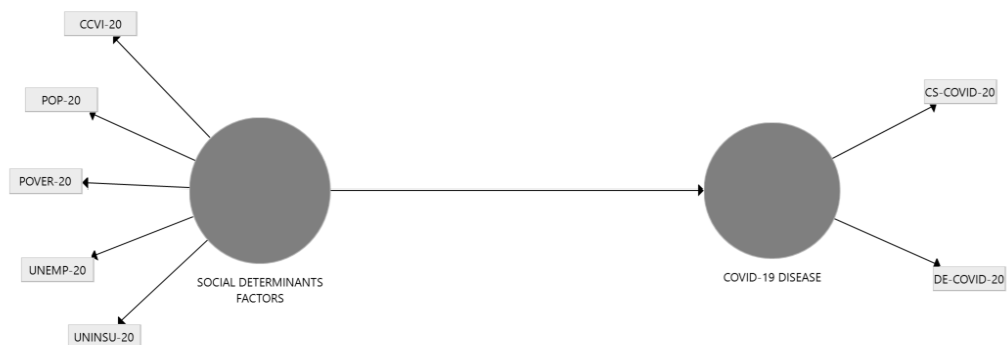

*Model evaluation*

Before starting with the respective multivariate analysis; Hair et al. (2012); Martínez & Fierro (2018a) establish the importance of their evaluation, which implies examining the reliability, internal consistency, convergent, and discriminant validation. These evaluations yielded the following results explained below.

Table 3. Precision and accuracy tests

| Code | Cronbach's α | AVE | rho_A | CR | VIF |
|---|---|---|---|---|---|
| SDF | .76 | .55 | .93 | .84 | 1.00 |
| C19-D | .88 | .89 | .91 | .94 | |

The results obtained show construct reliability in the model since the tests obtained values higher than p-value = .7. Regarding the convergent validation through the test (AVE), we conclude that the set of indicators represents a single underlying construct since values higher than p-value = .50 were obtained (Martínez & Fierro, 2018).

Therefore, each construct explains at least 50% of the variance of the indicators. When evaluating the collinearity level, the test (VIF) did not find problems related to collinearity since its values fluctuated at a p-value = 1.00. In the discriminant validity test or the Forner-Larcker Criterion, results in less than 0.7 confirm the existence of validity.

Table 4. Forner-Larcker Criterion

| Code | C19-D | SDF |
|---|---|---|
| C19-D | .94 | |
| SDF | .91 | .74 |

The model's predictive quality was performed using the Stone-Geisser redundancy test of cross-validation of the construct or Q2, which assesses the structural and theoretical model; with the results obtained with a value greater than zero 0, the conclusion is drawn existence of predictive validity and relevance of the model (Thaisaiyi, 2020).

Table 5. Constructs and indicators crossvalidated redundancy.

| Constructs | $Q^2$ | Indicators | $Q^2$ |
|---|---|---|---|
| C19-D | .72 | | |
| | | CS-COVID-20 | .85 |
| | | DE-COVID-20 | .58 |

Table 6. Constructs and indicators crossvalidated communality.

| Constructors | $Q^2$ | Indicators | $Q^2$ |
|---|---|---|---|
| SDF | .40 | | |
| C19-D | .56 | | |
| | | CCVI-20 | .46 |
| | | POVER-20 | .77 |
| | | POP-20 | .70 |
| | | UNEMP-20 | .06 |
| | | UNINSU-20 | .006 |

*Magnitude and significance of the model*

*Path coefficient results (β) and values (p)*

The analysis of the PLS algorithm's magnitude and significance allows us to measure and test the research model's respective hypothesis relationships. The magnitude is observed in the standardized regression coefficient (β) and its significance (p). With the Bootstrapping algorithm, the magnitude is observed in the standardized regression coefficient (β), and the significance in the two-tailed t (4900) values; where the critical value is (0.01; 4999) = 2,576 (Martínez & Fierro, 2018a). The resampling analysis evaluated (5000 subsamples) with a confidence level of 0.05.

Figure 2. Total effects SDF – C19-D ratio

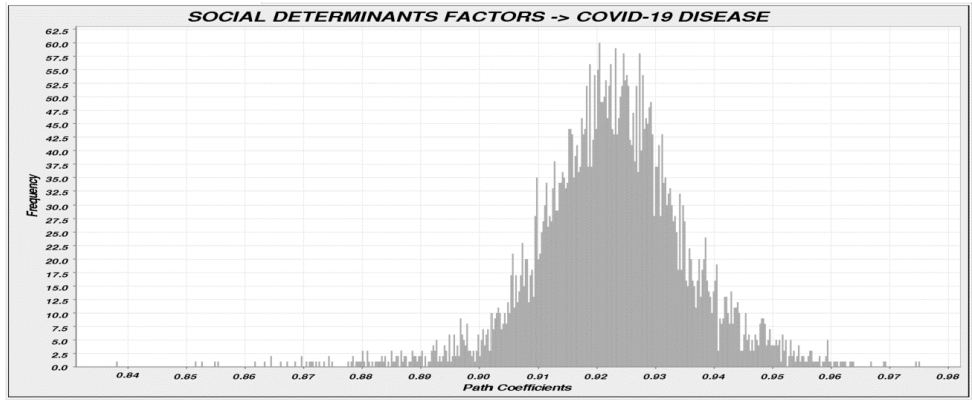

*Results*

The confirmatory analysis of the PLS least squares regression test establishes a high and robust significant impact between social determinants factors and covid-19 disease through a predictive value of R2 = .916, β = .836, p = .000 (t-value = 66.137). The research showed that for every unit of increase in social determinants, COVID-19 disease increases by 91.6%.

Table 7. Hypothesis test results

| Hypothetical Relationship | Coefficient β | t Student Boostrapping | $R^2$ | p | Hypothesis Accept |
|---|---|---|---|---|---|
| SDF-----C19-D | .836 | 66.137 | .916 | .000 | YES |

*Second analysis phase: correlational party-control analysis*

We will use the K-Means method to meet our grouping criteria (the axiom of non-negativity, identity, symmetry, and triangular inequality) of the data obtained in the two political groups and the silhouette analysis with the Manhattan metric to define the proximity of the study elements and identify the optimal value of k during the clustering process.

Figure 3. Political-control groups by state C1 (Republicans) and C2 (Democratic)

Figure 4. Party-control states lineal projection analysis

Note: It shows a two-dimensional projection with data from the Republican (red) and Democratic (blue) states, where the characteristics delineated by their respective vectors are combined.

The correlational modeling between social determinants and covid-19 disease also generated a new data set using the clustering index as a class attribute, which will allow the identification of patterns and the detection of clusters (Leong & Yue, 2017). It will also allow us to observe its constancy in time and meaning (Sabogal, 2013). The data are composed of three groups C1 Republicans, C2 Democrats, C3 California, New York, Texas, and Florida.

Figure 5. Political-control groups by state C1 (Republicans), C2 (Democratic) and C3 (TX, CA, NY & FL)

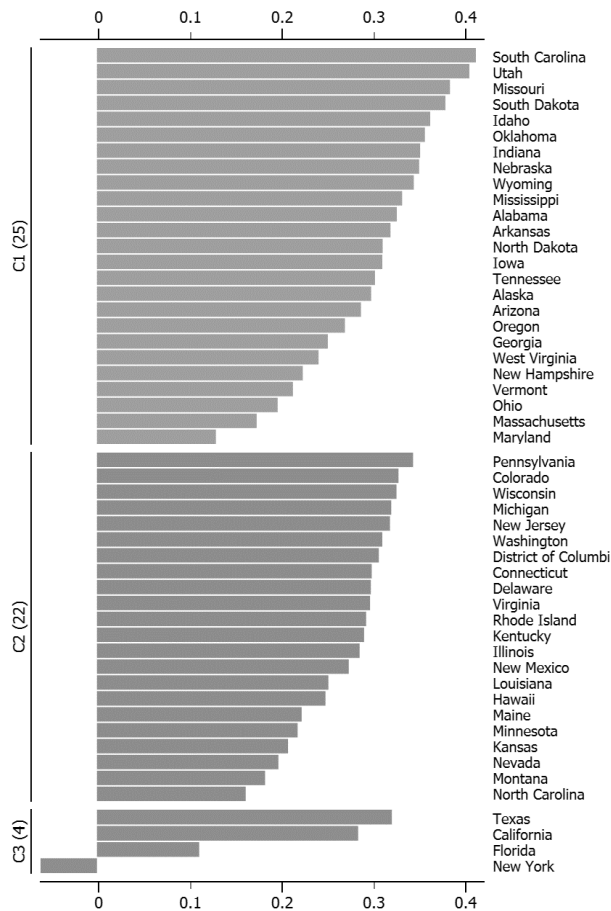

Figure 6. Clustering lineal projection analysis

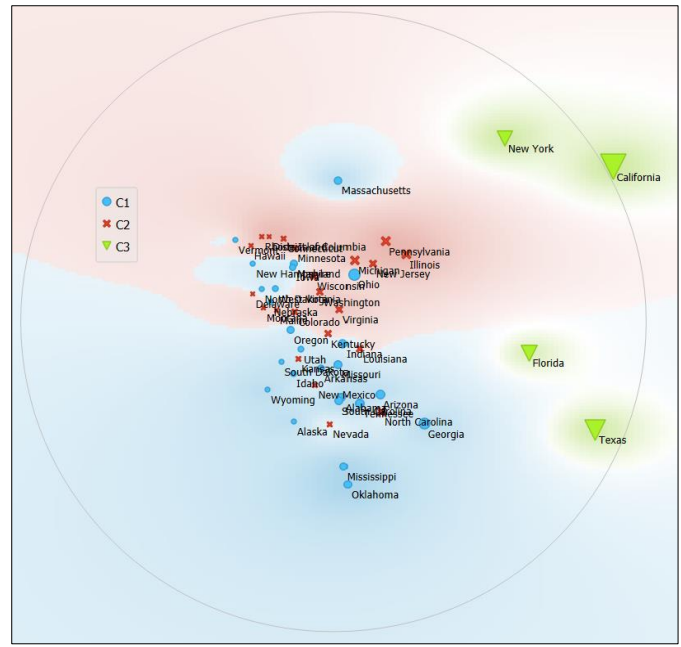

Note: Red represents the Republican Party, blue the Democratic Party, and the green represents CA, TX, FL, and NY

*Results*

The results obtained in the exploratory correlation analysis confirm the high relationship between the social determinants factors and the covid-19 disease in the following states under Republican and Democratic political control:

- High positive correlations found in California (D), Texas (R), Florida (R), and New York (D) belonging to group C3 between the number of people living below the poverty level and the number of infections registered by COVID-19 with ($r = .76$) and concerning with the number of deaths registered by COVID-19 ($r = -.53$).

- High positive correlations found in California (D), Texas (R), Florida (R), and New York (D) belonging to group C3 between the Covid-19 community vulnerability factor index and the number of infections registered by COVID-19 with ($r = .72$) and concerning with the number of deaths registered by COVID-19 ($r = -.80$).

- High positive correlations found in California (D), Texas (R), Florida (R), and New York (D) belonging to group C3 between the number of inhabitants and the number of infections registered by COVID-19 with ($r = .75$) and concerning with the number of deaths registered by COVID-19 ($r = -.55$).

- Moderate positive correlations found in California (D), Texas (R), Florida (R), and New York (D) belonging to group C3 between people without health insurance and the number of infections registered by COVID-19 with ($r = .66$) and concerning with the number of deaths registered by COVID-19 ($r = -.63$).

Figure 7. Party-control states group C3 correlation analysis

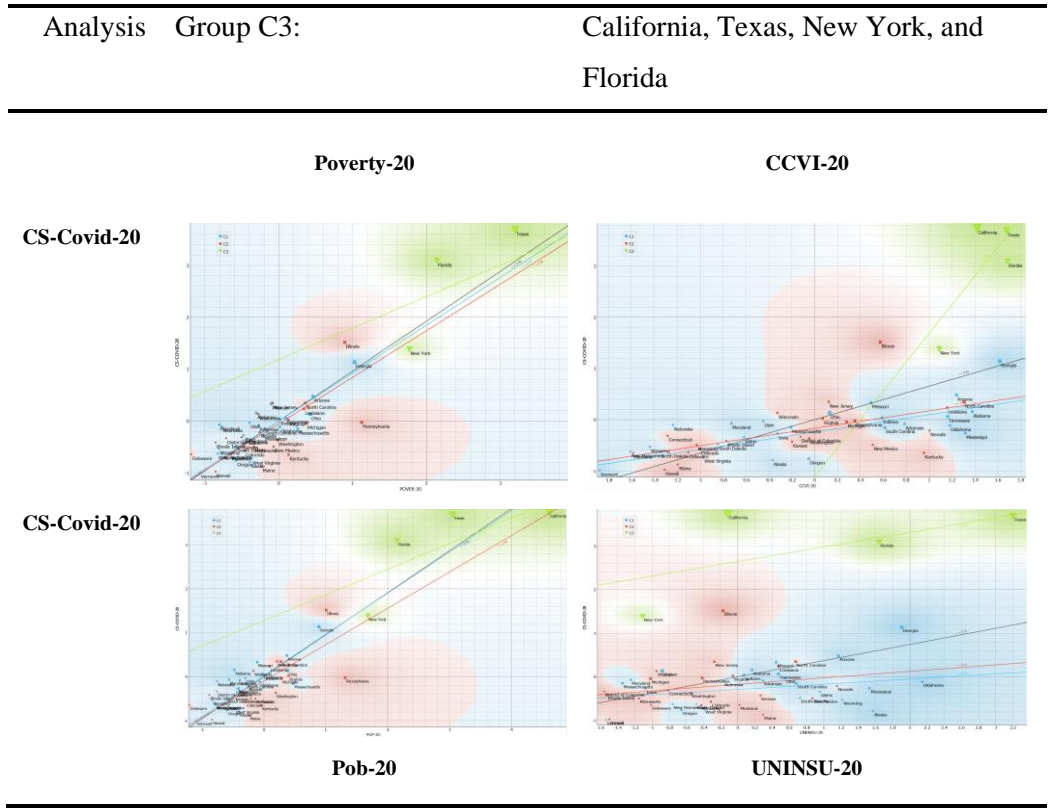

| Analysis | Group C3: | California, Texas, New York, and Florida |
|---|---|---|

*Republican Sates Correlation Analysis*

- High positive correlations found in all Republican States belonging to group C1 between the number of people living below the poverty level and the number of registered COVID-19 infections with ($r = .91$) and concerning with the number of deaths registered by COVID-19 ($r = .77$).
- High positive correlations found in all Republican States belonging to group C1 between the Covid-19 community vulnerability factor index and the number of COVID-19 infections registered with ($r = .80$) and concerning with the number of deaths registered by COVID-19 ($r = .54$).
- High positive correlations found in all Republican States belonging to group C1 between the number of inhabitants and the number of COVID-19 infections registered with ($r = .89$) and concerning with the number of deaths registered by COVID-19 ($r = .83$).
- Low positive correlations found in all Republican States belonging to group C1 between people without health insurance and the number of COVID-19 infections registered with ($r = .26$) and concerning with the number of deaths registered by COVID-19 ($r = -.08$).

*Democratic Sates Correlation Analysis*

- High positive correlations found in all Democratic States belonging to group C2 between the number of people living below the poverty level and the number of registered COVID-19 infections with ($r = .88$) and concerning with the number of deaths registered by COVID-19 ($r = .70$).
- Moderate positive correlations found in all Democratic States belonging to group C2 between the Covid-19 community vulnerability factor index and the number of COVID-19 infections registered with ($r = .59$) and concerning with the number of deaths registered by COVID-19 ($r = .39$).

- High positive correlations found in all Democratic States belonging to group C2 between the number of inhabitants and the number of COVID-19 infections registered with ($r = .89$) and concerning with the number of deaths registered by COVID-19 ($r = .74$).

- Low positive correlations found in all Democratic States belonging to group C2 between people without health insurance and the number of COVID-19 infections registered with ($r = .21$) and concerning with the number of deaths registered by COVID-19 ($r = .07$).

Figure 8. Party-control states group C3 correlation analysis

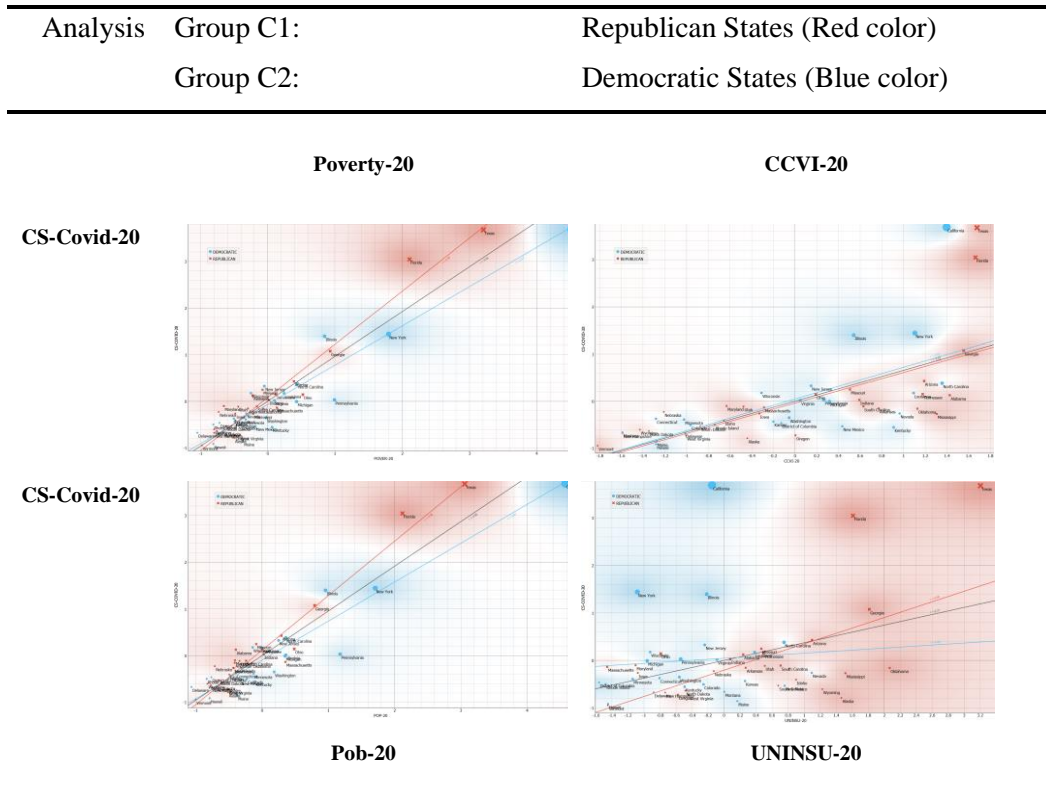

**Discussion**

Tables 4 and 5 indicate a high positive correlation between poverty levels and high contagion rates in states under Republican control or group C1, followed by states under Democratic control or group C2. The findings also show high values for death cases in New York compared to the other C3 states.

Figure 9. Time series graphic of deaths on group C3.

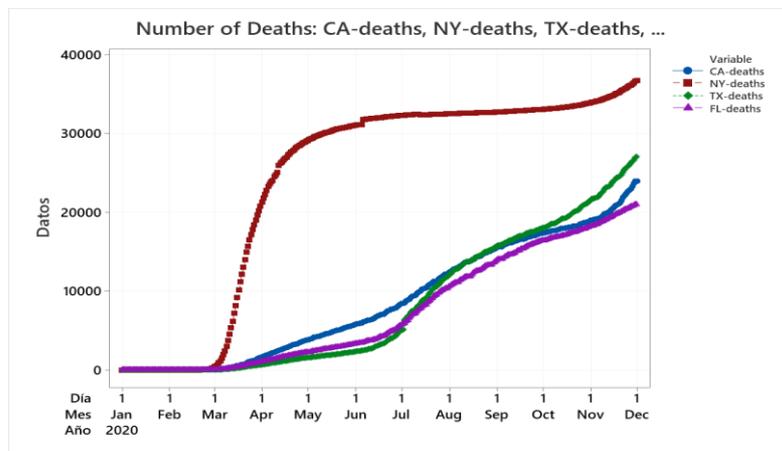

Note: Data on deaths from COVID-19 collected in different periods from January to December 2020.

There is emerging evidence that establishes that risk conditions linked to Poverty such as obesity, cardiovascular diseases, diabetes, and hypertension are risk factors for death from COVID-19; consequently, people with low socioeconomic status are more susceptible to mortality from infection (Patel et al., 2020).

Table 8. U.S. party-control correlation analysis

| PARTY-CONTROL STATES | POVER-20 vs CS-COVID-20/ DE-COVID-20 | CCIV-20 vs CS-COVID-20/ DE-COVID-20 | POP-20 vs CS-COVID-20/ DE-COVID-20 | UNINSU-20 vs CS-COVID-20/ DE-COVID-20 |
|---|---|---|---|---|
| REPUBLICAN | .98 \| .93 | .68 \| .62 | .98 \| .95 | .58 \| .42 |
| DEMOCRATIC | .97 \| .73 | .63 \| .53 | .97 \| .73 | .08 \| - .09 |

Figure 10. Poverty by political-control states 2020

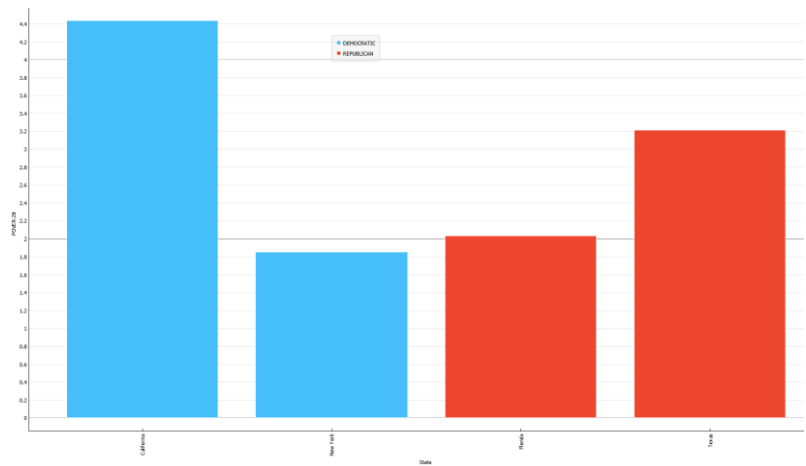

Note: Figures 10 and 11 compare the values of the poverty level between the four states of group C3 (CA, NY, XT, and FL).

Figure 11. Coverage without insurance health by political-control states 2020

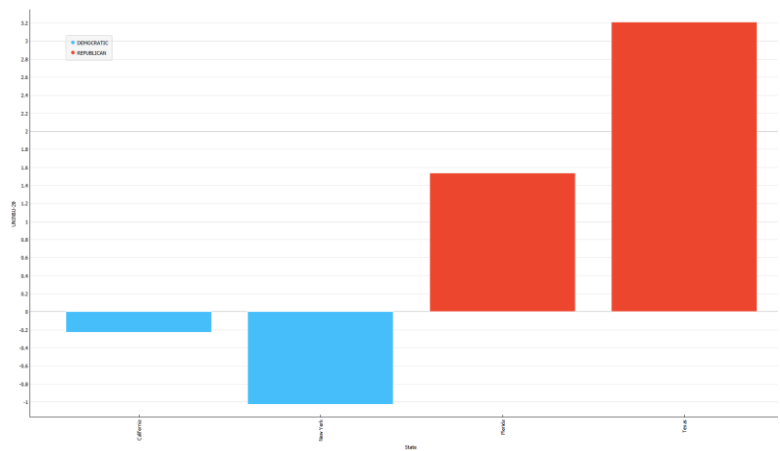

Table 9. U.S. Party-control regional correlation analysis

| Clusters | POVER-20 vs CS-COVID-20/ DE-COVID-20 | CCIV-20 vs CS-COVID-20/ DE-COVID-20 | POP-20 vs CS-COVID-20/ DE-COVID-20 | UNINSU-20 vs CS-COVID-20/ DE-COVID-20 |
|---|---|---|---|---|
| C1 - Republican | .91 \| .77 | .80 \| .54 | .89 \| .83 | .26 \| -.08 |
| C2 - Democratic | .88 \| .70 | .59 \| .39 | .89 \| .74 | .21 \| .07 |
| C3 – CA, NY, TX & FL | .76 \| -.53 | .72 \| -.80 | .75 \| -.55 | .66 \| -.63 |

The findings show a high positive correlation between the uninsured population with a health plan and high virus infection levels in California, Texas, New York, and Florida. Finally, the results show a high positive correlation between Poverty and high levels of COVID-19 infection than the other indicators that make up the Covid-19 Community Vulnerability Factors Index (CCIV), which explains the argument that Poverty and lack of economic security puts a public or private health system at risk and calamity (Weida et al., 2020).

**Conclusions**

In the last two presidential elections, the online strategies carried out by former presidents Obama and Trump became visible in what has been called social media elections (Shmargad & Sanchez, 2020). However, although both political parties maintained an active presence on social media in the last elections of 2020, a pattern of misinformation based on denial and conspiracy theories unleashed a lack of clear and reliable public health policies.

In the first spectrum, state governors who downplayed the Center for Disease Control and Prevention recommendations saw a disproportionate increase in infections and deaths (CDC, 2019). The evidence shows that the risk factor is the population with a lack of sufficient income to satisfy their basic needs. However, although unemployment grew dramatically, the evidence establishes that the unemployed population was not a risk factor. For this reason, it is necessary to deepen with more exploratory studies that identify and evaluate the causes of the high mortality rates that contrast with the poverty and coverage data of the medical plans of states such as New York.

On a second spectrum, the government's responsibility to address the factors that leave the most economically disadvantaged vulnerable to the virus, expanding the coverage of government health plans and actively contributing to minimizing social inequalities based on ethnic minority groups. The pandemic highlighted social and economic inequalities within American society and is likely to exacerbate them by considering more contagious variants, as there are high levels of transmission.

      Consequently, the executive and legislative branches' correct political decision-making is relevant in the framework of public health, addressing the vulnerabilities of the economically disadvantaged within American society with new, more inclusive health policies to help millions of American citizens living below the poverty line.